\documentclass[
	11pt,				
	a4paper			
	]{article}

\usepackage{lmodern}			
\usepackage[T1]{fontenc}		
\usepackage[utf8]{inputenc}	
\usepackage{indentfirst}		
\usepackage{nomencl} 			
\usepackage{color}				
\usepackage{graphicx}			
\usepackage{microtype} 			
\usepackage{amsmath, amsfonts}
\usepackage{subfig}
\usepackage{placeins}
\usepackage[sorting=none]{biblatex}

\title{Shearless bifurcations in particle transport for reversed shear tokamaks}

\author{G. C. Grime$^1$, M. Roberto$^2$, R. L. Viana$^{1,3}$\footnote{Corresponding author. e-mail: viana@fisica.ufpr.br}, Y. Elskens$^4$, I. L. Caldas$^1$ \\ 
\small 1.University of São Paulo, Institute of Physics, São Paulo, Brazil\\
\small 2. Aeronautics Institute of Technology, Physics Department, São José dos Campos, Brazil\\
\small 3. Federal University of Paraná, Physics Department, Curitiba, Brazil\\
\small 4. Aix-Marseille Université, UMR 7345 CNRS, PIIM, Campus Saint-Jér\^ome, Marseille, France
}

\definecolor{blue}{RGB}{41,5,195}

\makeatletter

\usepackage[margin=2.5cm]{geometry}

\begin{document}
\maketitle
	
\begin{abstract}
\noindent Some internal transport barriers in tokamaks have been related to the vicinity of extrema of the plasma equilibrium profiles. This effect is numerically investigated by considering the guiding-center trajectories of plasma particles undergoing ${\bf E}\times{\bf B}$ drift motion, considering that the electric field has a stationary nonmonotonic radial profile and an electrostatic fluctuation. In addition, the equilibrium configuration has a nonmonotonic safety factor profile. The numerical integration of the equations of motion yields a symplectic map with shearless barriers. By changing the parameters of the safety factor profile, the appearance, and breakup of these shearless curves are observed. The successive shearless curves breakup and recovering  is explained using concepts from bifurcation theory. We also present bifurcation sequences associated to the creation of multiple shearless curves. Physical consequences of scenarios with multiple shearless curves are discussed.

\

\noindent\textbf{Keywords:} drift motion, reversed shear profile, shearless curve, internal transport barrier
\end{abstract}

\section{Introduction}

The control of radial particle transport in tokamak plasmas is a necessary, albeit not sufficient, condition for obtaining good confinement, and it is currently an area of intensive research \cite{hazeltine,bookHB}. Such a goal can be achieved by creating internal transport barriers (ITBs), which are regions of reduced radial (cross field) particle transport in the plasma column \cite{horton99}. ITBs have been produced in JET by the utilization of strong supplementary heating during the current rise phase of the plasma discharge \cite{wolf}.

The most widely studied type of ITB is the so-called edge transport barrier (ETB), related to steep pressure gradients at the plasma edge \cite{wolf,garbet}. These gradients arise due to the high pedestal pressure profiles, characteristics of H-mode confinement, and can also exist inside the plasma core \cite{goldston,yushmanov}. This H-mode regime was obtained by combining neutral beam heating with a divertor configuration, showing a reduction of the particle and energy transport fluxes \cite{burrell,connor,wagner}.

Recently, a second type of ITB has been investigated, the shearless transport barriers (STBs), which appears in tokamak plasmas with reversed shear profiles, and for which the pressure gradients do not need necessarily to be as large as for ETBs \cite{shearless}. Reversed shear profiles can be obtained by modifying the safety factor profile \cite{levinton,strait,fujita1997}, and by applying radial electric fields in a specific way \cite{horton98}.  For example, reversed shear profiles of $q(r)$ have one or more extrema, at which shearless toroidal magnetic surfaces are formed \cite{morrison}. Shearless surfaces represent ITBs in the sense that cross-field transport is reduced therein \cite{wolf,gobbin2011}.

Another type of reversed shear appears by a suitable alteration of the tokamak equilibrium, what creates a nonmonotonic radial electric field profile \cite{horton98,marcus}. The presence of ITBs due to this effect can explain the reduction of turbulence-driven particle fluxes observed in tokamak experiments, leading to an improvement of the plasma confinement \cite{marcus2}. Finally, ITBs have been related to reversed shear profiles of the toroidal plasma velocity, measured in the Texas Helimak, where a set of probes is mounted to quantify velocity shear in different directions \cite{gentle,toufen}. 

The STBs are formed in mixed phase space with nonmonotonic equilibrium profiles, containing regular particle trajectories as invariant curves and chaotic trajectories \cite{diego1}. Locally, the invariant curves separate the chaotic trajectories and prevent the global chaotic particle transport in phase space \cite{diego2000}. Moreover, in STBs there are robust curves, namely \emph{shearless curves}, which survive to increasing chaotic area and, consequently, are among the last invariant curves to be broken.  This is a dynamical effect due to the perturbed trajectories in nonmonotonic plasma profiles \cite{diego1}. The onset of shearless transport barriers may be one of the mechanisms that hide behind certain improved scenarios observed in tokamaks with nonmonotonic plasma profiles. In this work, we apply a model proposed in Ref. \cite{horton98} to show how this mechanism depends on the required profiles.

The presence of different types of reversed shear (safety factor, radial electric field, toroidal plasma velocity) has been investigated by using a drift wave test particle transport model \cite{horton98}. The latter is based on ${\bf E}\times{\bf B}$ drift combined with the advection of guiding-center motion along the magnetic field lines \cite{horton98}. The numerical integration of guiding-center trajectories leads to a Poincaré stroboscopic map, in which we sample the coordinates at integer multiples of some characteristic period \cite{lichtenberg}. This kind of description, integrating particles trajectories, has led to important advances in the understanding of anomalous particle transport \cite{miskane,horton85,horton98,kwon2000,marcus,rosalem1,rosalem2,mouden}, trapped particle transport \cite{white1984} and generic magnetic perturbations  \cite{white2012,zestanakis2016}. 

The role of equilibrium profiles and oscillation spectrum in Horton's model have been studied previously. In Ref. \cite{rosalem1}, the influence of the nonmonotonic equilibrium electric field profiles on the nontwist transport barriers. Meanwhile, in the plasma edge, the trajectories are mainly determined by the plasma velocity profile \cite{rosalem2}. The effect of the amplitude of oscillation was also studied in Refs. \cite{mouden,marcus,osorio}. However, the influence of the safety factor profile has not yet been systematically investigated.

In this paper, we present a numerical investigation of the formation of single or multiple shearless curves due to reversed magnetic and electric shear profiles. The use of an $\mathbf{E}\times\mathbf{B}$ drift guiding-center description allows us to choose the radial profiles for the safety factor, radial electric field, and toroidal velocity. In this way, we are able to use reversed shear profiles, in order to study shearless transport barriers. A numerically obtained Poincaré map is used to compute the rotation number profile, which takes into account all reversed shear profiles. It turns out that there can be one or more STBs, corresponding to extrema of the rotation number profiles. Shearless curves can be created or destroyed by bifurcations triggered by suitable changing the safety factor profile. We also identify the dynamical mechanisms causing these shearless bifurcations. 

This paper is organized as follows: in Section 2, we present the drift guiding-center model to be used in the numerical simulations and the construction leading to the Poincaré map to be used in this work. Section 3 introduces the different reversed shear profiles to be considered in the numerical simulations, showing the appearance of shearless surfaces at the extrema of rotation number profiles. In Section 4, we discuss the possible shearless bifurcation scenarios. Our conclusions are left to the final Section.

\section{Drift-Wave Model}

One of the characteristic features of anomalous cross-field transport in tokamak plasmas is the presence of electrostatic wave instabilities arising from density and temperature gradients \cite{bookHB,horton1999drift}. A wide spectrum of waves has been shown to produce radial transport fluxes of plasma particles \cite{horton99}.  A mathematical model for describing the test particle motion in electrostatic waves has been proposed by \cite{horton98}, using a local coordinate system $\mathbf{x}=(r,\theta,\varphi)$, where $r$ is the radius measured from the magnetic axis, with $\theta$ and $\varphi$ being the poloidal and toroidal angles, respectively. We denote by $a$ and $R$, respectively, the minor and major plasma radius.

We consider the combined presence of an equilibrium magnetic field ${\bf B}$ and an electric field ${\bf E}$ related to the electrostatic waves. Moreover, let us suppose that the plasma particles are test particles, i.e., they are influenced by the external fields but do not affect them.  In the applied model \cite{horton98}, the $\nabla{B}$ and curvatures drifts are neglected, and therefore the trapped particles transport is not taken into account. Under these assumptions, the guiding-center motion has two components: (a) a passive advection along the magnetic field lines, with velocity $v_\parallel$, and (b) an ${\bf E}\times{\bf B}$ drift velocity, so that the guiding-center equation of motion is  
\begin{equation}
   \label{eqm}
    \dfrac{d \textbf{x}}{dt} = v_\parallel \dfrac{\textbf{B}}{B} + \dfrac{\textbf{E}\times\textbf{B}}{B^2}.
\end{equation}

In the large aspect ratio approximation ($\epsilon = a/R \ll 1$), we use a tokamak equilibrium magnetic field $\mathbf{B} = \left(0,B_\theta(r),B_\varphi\right)$, where $B_\varphi$ and $B_\theta$ are the toroidal and poloidal magnetic field components, respectively. Since $B_\theta \sim \epsilon B_\varphi$ we approximate $B \approx B_\varphi \gg B_\theta$ and treat $B$ as a uniform field. The safety factor of the magnetic surfaces is thus given by \cite{hazeltine}
\begin{equation}
\label{qr}
    q(r) = \dfrac{r B_{\varphi}}{R B_\theta(r)}.
\end{equation}

The model applied in this paper enables us to relate the advanced scenarios in tokamaks with the STBs predicted for the nonmonotonic plasma profiles, typical of these scenarios. To investigate this relation, it is assumed in this model \cite{horton98} that the electrostatic fluctuation spectrum is coherent and time independent.

The assumed electric field can be expressed as ${\bf E} = {\overline{{\bf E}}_r(r)} - \nabla{\tilde\phi}$, where ${\overline{{\bf E}}_r}$ is a time-independent radial electric field profile, and $\tilde{\mathbf{E}} = - \nabla \tilde{\phi}$ is a fluctuating part, representing the electrostatic instabilities in the tokamak edge \cite{horton98}. The latter is supposed to exhibit a broad spectrum of frequencies $\omega_n = n \omega_0$ and wave vectors, characterized by the oscillation spectrum \cite{horton98}
\begin{equation}
\label{eq:spectrum}
\tilde{\phi}(\textbf{x},t) = \sum_{m,\ell,n} \phi_{m,\ell,n} \cos{(m\theta-\ell\varphi} -n\omega_0 t + \alpha_n),
\end{equation}
where $\alpha_n$ is the relative phase and $\phi_{m,\ell,n}$ are constant coefficients. A radial dependent spectrum was proposed in Ref. \cite{connor1987} and results in a \emph{global drift wave map}, valid for long times \cite{kwon2000}. However, to study local transport, we will use the \emph{local drift wave map}, valid for short times \cite{horton98}. In this local map, we assume the coefficients $\phi_{m,\ell,n}$ to be constants and consider only the dominant spatial mode in Eq (\ref{eq:spectrum}), with harmonics of the lowest frequency $\omega_0$, and poloidal and toroidal mode numbers $m = M$ and $\ell = L$, respectively. Although the electrostatic fluctuations have a broad spectrum, the wavenumber width is smaller than the frequency width \cite{marcus2}. Therefore, the numbers $M$ and $L$ can be estimated by the highest amplitudes in the fluctuation spectrum. In this case, the electrostatic fluctuation spectrum becomes
\begin{equation}\label{eq:total.spectrum}
\tilde{\phi}(\textbf{x},t) = \sum_{n}\phi_{n}\cos{(M\theta-L\varphi} -n\omega_0 t + \alpha_n).
\end{equation}

This model does not take into account the nonlinear interactions between the modes and the incoherent nature of the fluctuations. It supposes a single spatial mode of oscillation, and a temporal spectrum concentrated in low frequencies. The correlation time and length of the fluctuation spectrum is a key parameter for the validity of this theoretical analysis. If the correlation time $\tau_c$ is small compared to circumnavigation time of the $\mathbf{E}\times \mathbf{B}$ motion  $\tau_{circ}$, then the quasi-linear theory holds. However, if the perturbation has a correlation time long enough for a test particle to be able to feel the whole wave structure, we enter in trapping regime \cite{diamond}. The dimensionless Kubo number, defined by $K=\tau_c/\tau_{circ}$, gives us a simple criterion: in the limit $K\ll 1$ the quasilinear theory holds, and for $K>1$ the particle is in the trapping regime \cite{zimbardo2000}.

Writing Eq. (\ref{eqm}) in components, and taking into account the large aspect ratio approximation, there results
\begin{align}
    \label{eqr}
    \frac{dr}{dt} & = - \frac{M}{B r}  \sum_{n}\phi_{n}\sin{(M\theta-L\varphi} -n\omega_0 t + \alpha_n), \\
    \label{eqt}
    r \frac{d\theta}{dt} & = \frac{r v_\parallel(r)}{R q(r)} - \frac{\overline{E_r}(r)}{B}, \\
    \label{eqf}
    R \frac{d\varphi}{dt} & = v_\parallel(r).
\end{align}

The guiding-center equations of motion \eqref{eqr}-\eqref{eqf} are written in terms of the three radial profiles to be considered in this work. Defining action and angle variables \cite{horton98}, $I=(r/a)^2$ and $\psi = M\theta - L\varphi$, and performing a normalization (Appendix \ref{appendixA}), the set of three equations reduces to just two:
\begin{align}
\label{eq:model.adm}
\dfrac{dI}{dt} &= 2M\sum_n \phi_n\sin(\psi - n\omega_o t + \alpha_n),\\
\label{baba}
\dfrac{d\psi}{dt} &= \epsilon\dfrac{v_\parallel(I)}{q(I)} \left[ M-Lq(I) \right] - \dfrac{M{\overline{E_r}(I)}}{\sqrt{I}}.
\end{align}

The drift motion of guiding-center of charged particles in combined magnetic and electric fields has been long known to be a Hamiltonian system, with canonical equations \cite{morrison}
\begin{equation}
    \label{hh}
    \frac{dI}{dt} = - \frac{\partial H}{\partial\psi}, \qquad 
    \frac{d\psi}{dt} = \frac{\partial H}{\partial I},  
\end{equation}
where the Hamiltonian can be written as $H(I,\psi,t) = H_0(I) + H_1(\psi,t)$, where 
\begin{equation}
    \label{h0}
    H_0(I) = \int^I dI' \left\{
    \epsilon\dfrac{v_\parallel(I')}{q(I')} \left[ M-Lq(I') \right] - \dfrac{M{\overline{E_r}(I')}}{\sqrt{I'}}
    \right\}
\end{equation}
\noindent is the equilibrium part, and 
\begin{equation}
    \label{h1}
    H_1(\psi,t) = 2M\sum_n \phi_n \cos(\psi - n\omega_o t + \alpha_n)
\end{equation}
\noindent corresponds to the time-dependent perturbation. Hence, in general, the system is non-integrable. 

If we switch off the perturbation caused by electrostatic fluctuation (this amounts to set $\phi_n = 0$ for all values of $n$), Eq. (\ref{eq:model.adm}) shows that the action variable is a constant of motion, as required from an integrable system \cite{lichtenberg}. The perturbations of the quasi-integrable system \eqref{eq:model.adm}-\eqref{baba} have resonant and non-resonant modes. The resonance condition for a perturbation mode $n$ is given by $(d/dt)(\psi - n\omega_0 t) \approx 0$, which implies
\begin{equation}
\label{eq:resonant.condition}
    n\omega_0 = \epsilon\dfrac{v_\parallel(I)}{q(I)}[M-Lq(I)] - \dfrac{M \overline{E_r}(I)}{\sqrt{I}}.
\end{equation}

Once the profiles of $q(I)$, ${\overline{E}_r(I)}$ and $v_\parallel(I)$ are specified, it turns out that Eq. (\ref{eq:resonant.condition}) is satisfied for a given $n$ only for certain values of the action $I = I_n$. From Hamiltonian system theory follows that chains of periodic islands appear due to the perturbation, centred at those resonant values $I_n$. 

In order to visualize those islands, we can use a Poincaré map obtained stroboscopically, i.e., we sample the values of the action-angle variables at integer multiples of a characteristic period, which is $\tau = 2\pi/\omega_0$ in our system. Numerically integrating equations \eqref{eq:model.adm} we obtain stroboscopic Poincaré maps by plotting the trajectories in instants $t_j = j (2\pi /\omega_0)$. The two-dimensional map $I_{j+1} = I_{j+1}(I_j,\psi_j)$, $\psi_{j+1} = \psi_{j+1}(I_j,\psi_j)$ is area-preserving thanks to Liouville theorem \cite{hazeltine}. 

Since we are interested chiefly in nonmonotonic radial profiles corresponding to reversed shear quantities like $q(I)$, $v_\parallel(I)$, and $E_r(I)$, the Poincaré map is nontwist in general. Many results of Hamiltonian theory, like KAM theory and Aubry-Mather theory, hold only for twist systems, hence new features are expected to happen when nontwist maps are considered \cite{morrison,diego2}. One of them is the existence of twin island chains \cite{morrison,diego2}. These twin chains are centred at resonant values $I_n$ satisfying (\ref{eq:resonant.condition}) that arise in pairs. The evolution of map orbits near twin chains has been extensively investigated for the so-called standard nontwist map. 

Introducing reversed shear profiles in this model, transport barriers correspond to shearless invariant curves in the phase space, defined by an extreme point in rotation number profile \cite{morrison,diego1}. To every regular (nonchaotic) orbit we can associate a rotation number $\Omega$ given by the mean rotation angle in the Poincaré section. Given an initial condition $(I_0,\psi_0)$, the rotation number of this orbit is given by
\begin{equation}
\label{rota}
    \Omega (I_0) = \lim_{N\to \infty} \dfrac{\psi_N - \psi_0}{N},
\end{equation}
\noindent where $\psi_N$ is the angle of $N$-th intersection in the Poincaré map. We choose $\psi_0 = 0$ in numerical simulations, but the final value does not depend on that choice. The limit in Eq. (\ref{rota}) exists provided the orbit is non-chaotic.

\section{Breakup and reappearance  of shearless transport barriers}

{In this section, we show the presence of shearless invariant curves locally separating the chaotic trajectories and avoiding the plasma edge transport.}

In order to numerically solve equations \eqref{eq:model.adm}-\eqref{baba} we use  parameters of TCABR \cite{nascimento}, although the results can be applied to any tokamak, described in a large aspect ratio approximation. However, we present a conceptual investigation rather than detailed comparisons with experiments performed in any tokamak.

The improvement of plasma confinement quality by using reversed shear profiles has long been acknowledged. A substantial reduction of the turbulent transport levels has been observed in regions with negative magnetic shear \cite{mazzucato,levinton,strait,fujita1997}. In order to generate negative shear regions,  it is necessary that the safety factor radial profile be nonmonotonic. MHD-based models of a cylindrical plasma column suggest the following profile of the safety factor \cite{kerner1982tearing}:
\begin{equation}\label{eq:non.q.profile}
q(r)=q_a\dfrac{r^2}{a^2}\left[ 1- \left(1+\mu^{'}\dfrac{r^2}{a^2}\right)\left(1-\dfrac{r^2}{a^2}\right)^{\nu +1} \right]^{-1},
\end{equation}
where $q_a$ is the safety factor at plasma edge, and 
\begin{equation}
    \label{betaeq}
\mu ' = \mu \dfrac{\nu +1}{\mu + \nu + 2}.
\end{equation}

\begin{figure}
\centering
\includegraphics[width=0.8\textwidth]{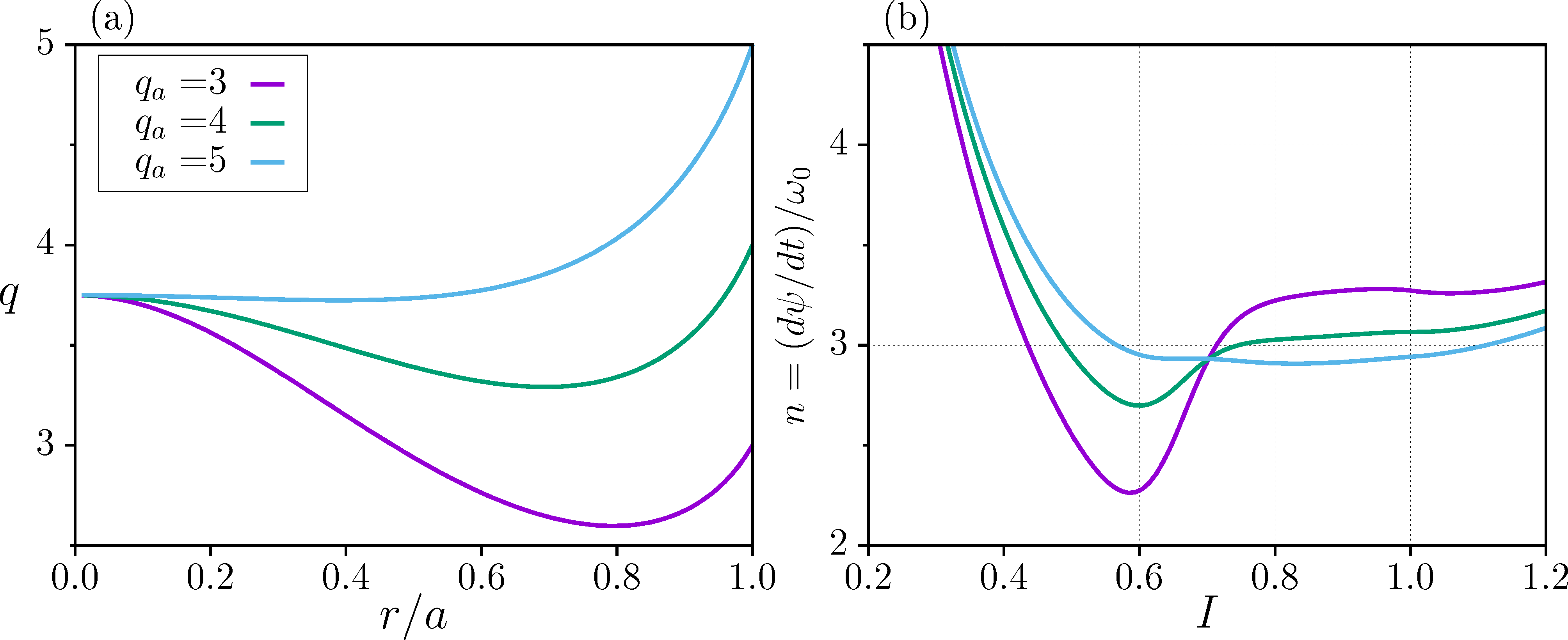}
\caption{\label{fig:q.profile}(a) 
Nonmonotonic safety factor profile for some values of parameter $q_a$. (b) Resonant mode numbers as a function of the action value for different values of $q_a$. Since only integer values of $n$ are allowed, there follows that only the modes $n = 3$ and $4$ produce resonances.}
\end{figure}

In the numerical simulations to be shown in this work, we fixed the parameters $\nu=0.8$ and $q_0=3.75$, making the remaining parameter $\mu$ a function of $q_a$, which we choose as our control parameter. The $q(r)$ profile is plotted on Figure \ref{fig:q.profile}(a) for some values of the control parameter. For $q_a = 3$ and $4$ the profile is nonmonotonic, with minima at $r/a \approx 0.7$ and $0.8$, respectively; whereas $q_a = 5$ yields a quasi-monotonic profile, included here for completeness. This range of values of $q_a$ is compatible with TCABR plasma discharges \cite{nascimento}.

The presence of negative shear in the radial electric field profile has been related to the reduction of turbulent particle fluxes in H-mode tokamak discharges \cite{viezzer,hidalgo}. This effect is compatible with a shearless transport barrier if the radial profile of $E_r$ is nonmonotonic \cite{marcus,marcus2}. One of the simplest functions with this property is a quadratic one, given by 
\begin{equation}
\label{eq:electric.field.profile}
\overline{E_r}(r) = 3\alpha r^2 + 2\beta r + \gamma, 
\end{equation}
where $\alpha = -1.14$, $\beta = 2.529$, and $\gamma = -2.639$ are parameter values after the normalization (see the Appendix for details). These values were chosen to yield a local minimum in the desired plasma region and are compatible with profiles measured in TCABR tokamak \cite{nascimento}. 

Toroidal plasma rotation is an important effect to be taken into account in tokamaks, such as stabilization or growth rate decrease of certain MHD modes. Besides intrinsic rotation, this effect can also be obtained by using neutral beam injection, since the incident particles impart momentum to the plasma particles. Moreover, it has been observed that plasma rotation is able to decrease turbulent flux levels in the plasma edge \cite{gentle,toufen}. Hence, it has been conjectured that nonmonotonic profiles of the toroidal plasma velocity can be related to internal transport barriers (actually, shearless barriers). 

Spectroscopic techniques have been used for the measurement of toroidal plasma rotation velocities in TCABR discharges, giving values about $3.0~\mathrm{km/s}$ for the plasma edge \cite{nascimento}. A normalized parallel velocity profile to be used in this work, and consistent with TCABR observations, is given by
\begin{equation}
\label{eq:toroidal.velocity}
v_\parallel(I) = v_{\parallel 0} + v_m \tanh{(\sigma_1 I + \sigma_2)}
\end{equation}
where the parameters take on the following values: $v_{\parallel 0} =-3.15$, $v_m=5.58$, $\sigma_1 = 14.1$, and $\sigma_2 = -9.26$, once we apply the normalization factor $v_0=E_0/B_0$ \cite{osorio}. 

After discussing the equilibrium aspects of the model, let us consider the perturbation caused by the electrostatic fluctuations given by Eq. (\ref{eq:total.spectrum}). We assume the spatial dominant mode to have $M/L=16/4$, which are typical numbers in the wave spectrum at the tokamak plasma edge \cite{horton98}. The temporal modes considered are $n=2,3,4$, based on the fluctuating spectrum of TCABR \cite{marcus2}, with normalized amplitudes $\phi_2=11.74 \times 10^{-3}$, $\phi_3=2.077 \times 10^{-3}$ and $\phi_4=0.2443 \times 10^{-3}$. The fundamental frequency of the temporal modes is around $10~\mathrm{kHz}$ \cite{marcus}, which implies a normalized angular frequency $\omega_0=5.224$. We shall keep $\alpha_n=0$ and the perturbation amplitudes at these values throughout this work.

Assuming that the rest of profiles and parameters are fixed, the safety factor will be chosen to be the tunable parameter which determines the dynamical behaviour of the system. Given a $q_a$ value, as mentioned before, there are resonant actions $I_n$ determined by the condition \eqref{eq:resonant.condition}, whose numerical solution is shown in Figure \ref{fig:q.profile}(b) for different values of $q_a$. The mode $n=3$ produces a resonance at two values of the action $I_{3,(a,b)}$, which also  characterizes nontwist behaviour. The $n=4$, however, gives a resonance at a  single value $I_4$. In addition, the $n=2$ mode does not yield resonance at any value of $I$, but it nevertheless influences in the formation and destruction of STBs \cite{marcus}.

\begin{figure}
\centering
\includegraphics[width=0.9\textwidth]{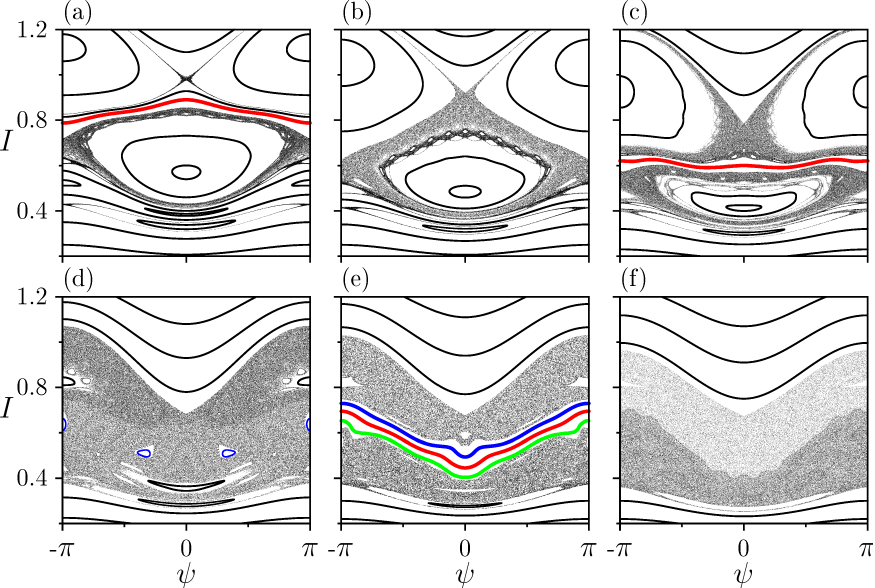}
\caption{\label{fig:poincare.general} Poincaré sections in the action-angle variables obtained by numerical integration of the equations of drift motion, for different values of the safety factor at plasma edge $q_a$: (a) $q_a = 5.00$, (b) $q_a = 4.50$, (c) $q_a = 4.00$, (d) $q_a = 3.45$, (e) $q_a = 3.30$, and (f) $q_a = 3.00$. The shearless curves are indicated by green, red, and blue wherever they appear in the Poincaré sections.}
\end{figure}

The variation of the safety factor profile changes the behaviour of shearless transport barriers. Figure \ref{fig:poincare.general} displays  examples of Poincaré sections for some values of control parameter $q_a$. We represent the action-angle variables as rectangular coordinates for better visualization. In Figure \ref{fig:poincare.general}(a), obtained for $q_a = 5.00$, we observe two large (twin) islands and a chaotic region around the inner one (the outer island has also a chaotic region, but it is too narrow to be seen with the present resolution). 

\begin{figure}
    \centering
    \includegraphics[width=0.8\textwidth]{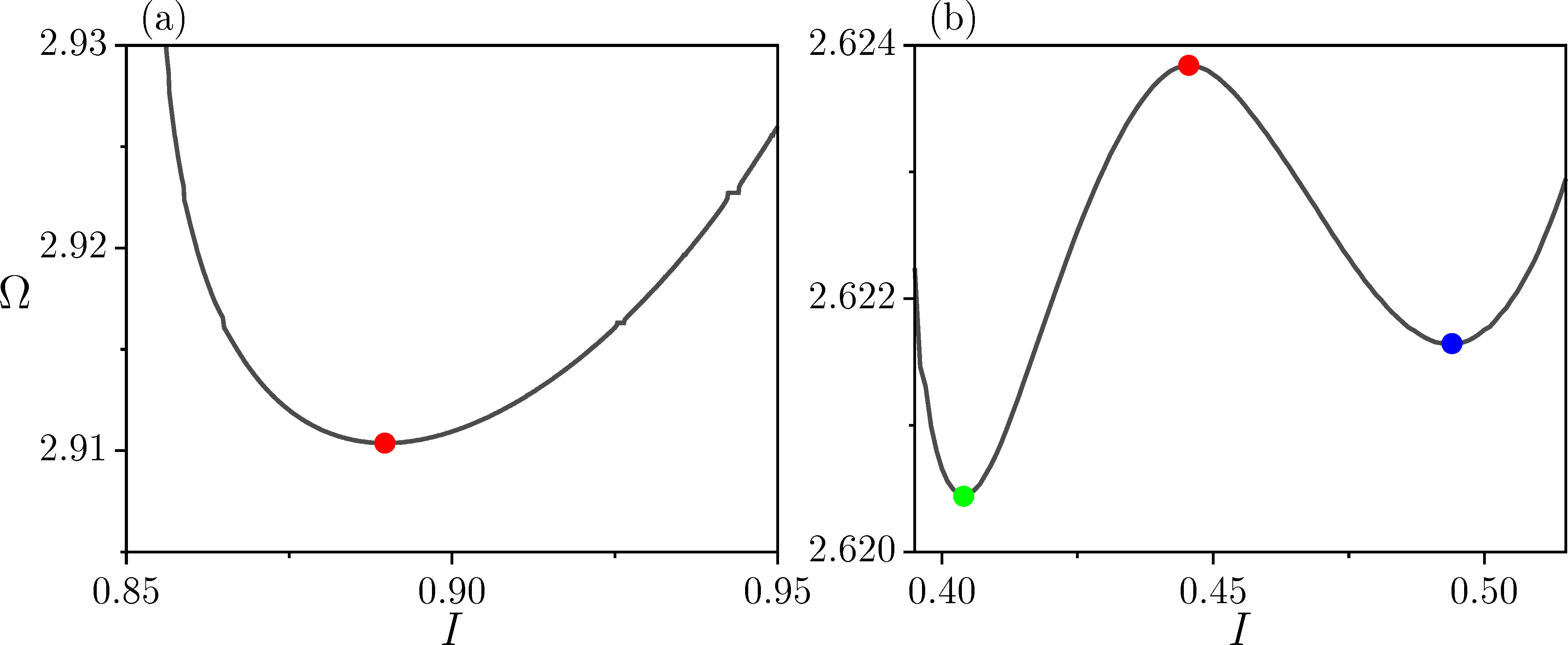}
    \caption{Rotation number profile corresponding to the Poincaré sections depicted in Figures \ref{fig:poincare.general}(a) and (e). 
    The extrema for each case are indicated by red, green, and blue points, in the case of one, two, and three coexisting shearless curves, respectively.}
    \label{fig:rotation.shearless}
\end{figure}

Those islands refer to the main resonances of $n=3$ mode in Figure \ref{fig:q.profile} and are direct consequences of the nonmonotonicity of the profiles. Between these twin islands there is a shearless curve,  located at the action value corresponding to an extremum of the rotation number profile  [Fig. \ref{fig:rotation.shearless}(a)]. There are other island chains corresponding to higher order resonances, but their width is considerably smaller than the main ones, and their effect will not be taken into account, at least directly.

The chaotic region around the inner islands grows as the parameter $q_a$ decreases and eventually causes the breakup of the shearless curve for $q_a = 4.50$ [Figure \ref{fig:poincare.general}(b)]. The mechanism of shearless curve breakup due to increasing perturbation has been thoroughly described by Morrison and co-workers, in the context of the standard nontwist map (SNM), introduced by Del Castillo and Morrison \cite{diego2,diego1}
\begin{align}
    \label{snm1}
    I_{n+1} & = I_n - {\tilde b} \sin(2 \pi \theta_n), \\
    \label{snm2}
    \theta_{n+1} & = \theta_n + {\tilde a} (1 - I_{n+1}^2),
\end{align}
where $(I_n,\theta_n)$ can be regarded as action-angle variables of the drift model, when we consider a quadratic approximation of the radial safety factor profile about a local extremum \cite{horton98}, and ${\tilde a}$, ${\tilde b}$ are system parameters. Moreover, a map (discrete time) model can be obtained if we consider a wide frequency spectrum of equally spaced resonant modes, instead of a single dominant resonant mode.

Noteworthy, if the value of $q_a$ is further decreased to $4.00$, the shearless curve between the two twin islands reappears [Fig. \ref{fig:poincare.general}(c)], since the corresponding rotation number profile has an extremum for this parameter value. This new shearless curve, on its way, is broken for smaller values of $q_a$, like $3.45$ [Fig. \ref{fig:poincare.general}(d)].

Further decrease in $q_a$ causes the appearance of three shearless curves at $q_a = 3.30$ [plotted in red, green, and blue in Fig. \ref{fig:poincare.general}(e)]. They are also located at extrema of the rotation number profile: the red one at a maximum, and the green and blue ones at minima [Fig. \ref{fig:rotation.shearless}(b)]. These multiple  shearless curves also disappear for even lower values of $q_a$, as exemplified by Fig. \ref{fig:poincare.general}(f), obtained for $q_a = 3.00$

\begin{figure}
\centering
\includegraphics[width=.4\textwidth]{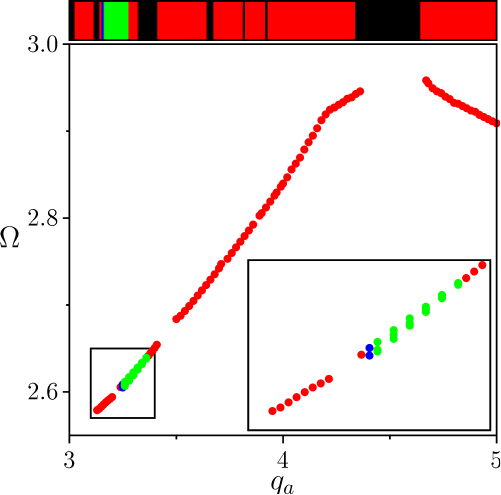}
\caption{\label{fig:shearless.bifurcation} Shearless bifurcation diagram showing the extrema of the rotation number profile (whenever they exist) as a function of the control parameter $q_a$. Blue and green points indicate the observation of two and three extrema, respectively. On the top colorbar, black regions represent intervals for which there is no shearless transport.}
\end{figure}

Sequences of breakup and reappearance of shearless curves as a parameter is varied, exemplified by Figs. \ref{fig:poincare.general}(a)-\ref{fig:poincare.general}(c), occur quite often in the nontwist system considered in this work. In Figure \ref{fig:shearless.bifurcation} we plot a shearless bifurcation diagram, showing the extrema of rotation number as a function of the parameter $q_a$. Hence, each point in this diagram corresponds to a shearless curve, whenever it exists. There are some gaps in the diagram, corresponding to parameter values with no extrema of $\Omega$, or for which the rotation number itself is ill-defined (because the considered orbit is chaotic, for example). Such regions are indicated in black colour in the bar above Fig. \ref{fig:shearless.bifurcation}

For most parameter values in the shearless bifurcation diagram, there is only one shearless curve, represented by red points in Fig. \ref{fig:shearless.bifurcation}, corresponding to the red regions in the colorbar. Multiple shearless curves are indicated by blue and green points, when there are two and three coexisting shearless curves, respectively. In an inset of Fig. \ref{fig:shearless.bifurcation} we display a magnification of the interval $3.1<q_a<3.4$, for which there are green points indicating three coexisting shearless curves. There are some blue points as well, but they occupy a region so tiny that they are barely visible in Fig. \ref{fig:shearless.bifurcation}. 

We see that a shearless bifurcation is an abrupt change in the behaviour of the shearless curve, as a system parameter is varied through a critical value. For example, a change between red and green points in Fig. \ref{fig:shearless.bifurcation} suggests the occurrence of a shearless bifurcation, whereby a single shearless curve bifurcates into two shearless curves, as the parameter $q_a$ passes through some value. When green points become red points again, we can speak of an inverse shearless bifurcation. Bifurcations on transport barriers like the one mentioned above were experimentally observed on JET and ASDEX Upgrade \cite{joffrin2003}.  However, in these experiments, the triggering mechanism of those barriers is different from the one presented in this paper, but the possibility to have more than one transport barriers remains.

\section{Shearless bifurcations scenarios}

The shearless curves undergo bifurcations of many types. In the most frequent, the shearless curve simply disappears, as a control parameter passes through some critical value. Since shearless curves represent transport barriers, their breakup will be followed by a significant increase in the transport levels \cite{szezech}. After the breakup of a shearless curve, a larger chaotic region would enable the guiding centres to undergo longer excursions in action space \cite{szezech}. However, if limited regions of chaotic orbits exist before the shearless curve breaks down, the curve break-up is not expected to increase significantly the transport levels.

\begin{figure}
    \centering
    \includegraphics[width=.9\textwidth]{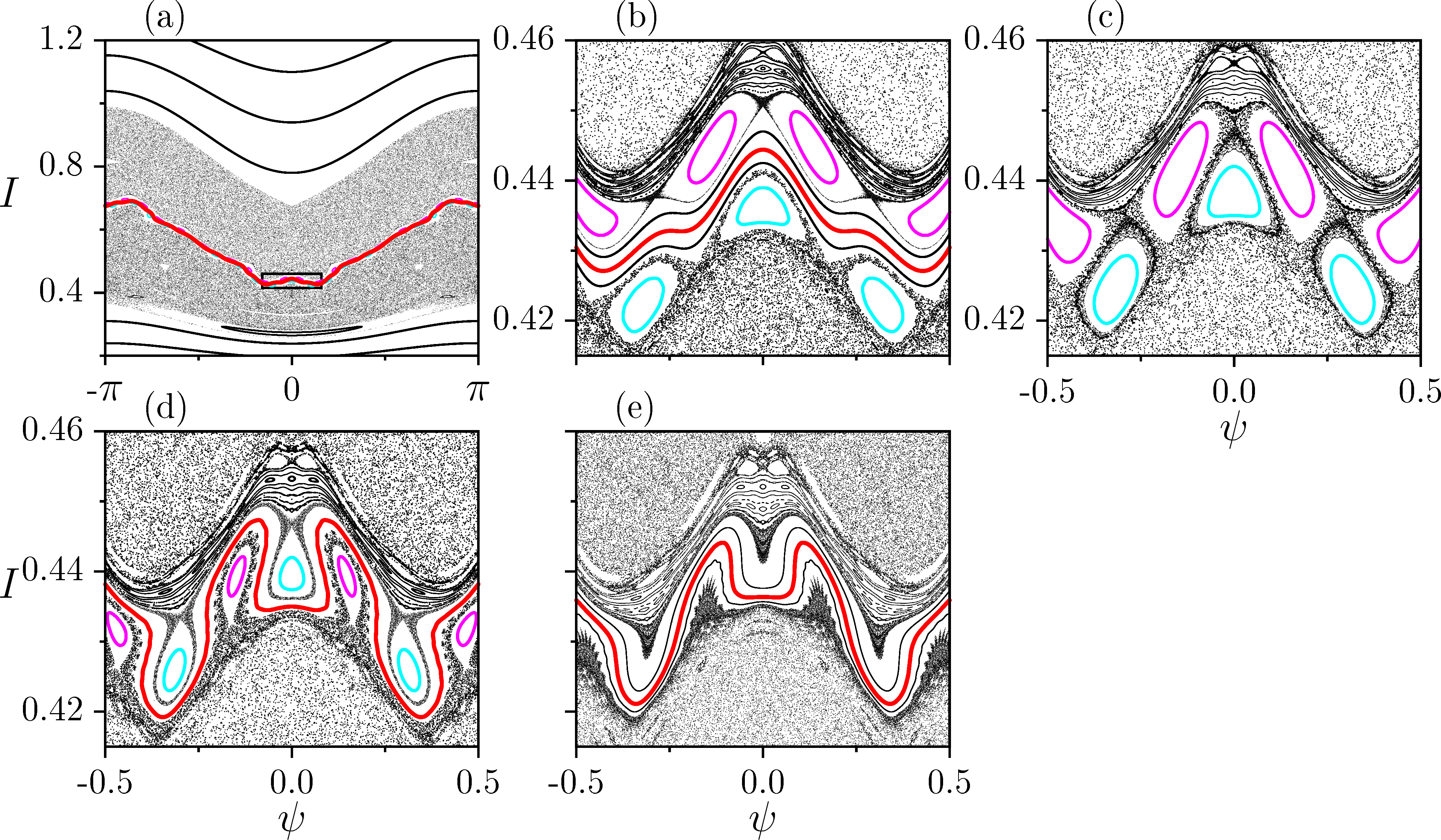}
    \caption{Poincaré sections of the drift wave model for {(a) $q_a=3.1290$, (b) magnification of marked region in (a), (c) $q_a=3.1282$, (d) $q_a=3.1277$ and (e) $q_a=3.1272$}. The remaining parameter values are the same as in Figure 2. In these phase portraits, we magnify the region containing the twin islands of $19$ islands each, straddling the shearless curve.}
    \label{fig:odd.scenario}
\end{figure}

\begin{figure}
\centering
\includegraphics[width=0.8\textwidth]{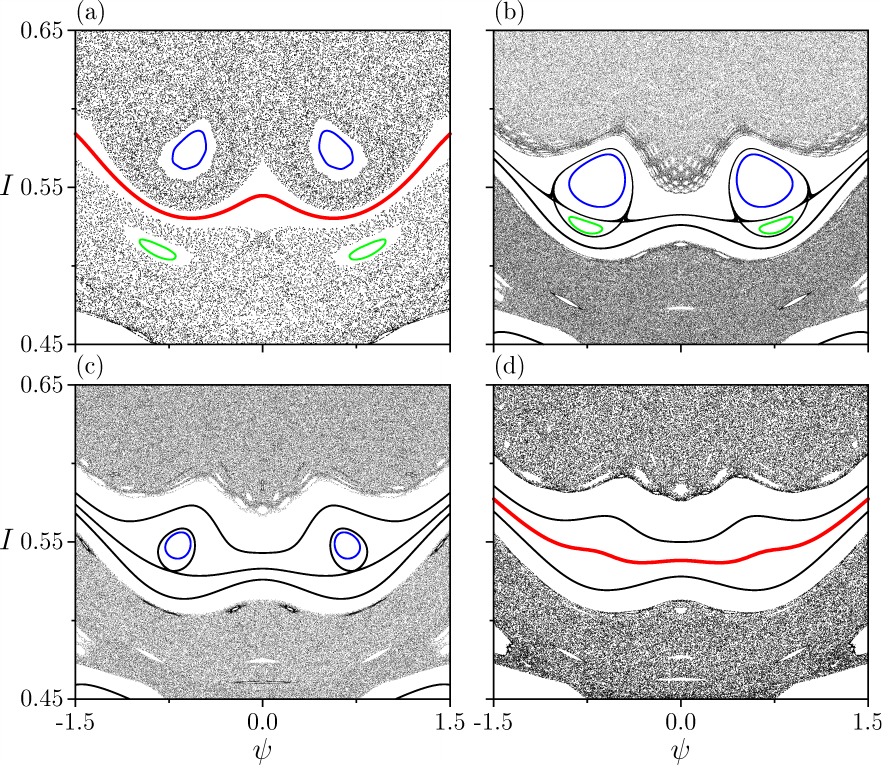}
\caption{\label{fig:even.scenario} Poincaré sections of the drift wave model for  (a)  $q_a=3.700$ (b) $q_a=3.725$ (c) $q_a=3.730$ and (d) $q_a=3.740$. The remaining parameter values are the same as in Figure 2.}
\end{figure}

In the SNM, after the shearless curve disappears, global chaotic transport appears in phase space, since at both sides of the shearless curve there are locally chaotic regions that merge together after the curve breakup, leaving a larger chaotic region therein. We observed that the system \eqref{eq:model.adm}-\eqref{baba} manifests some atypical shearless breakups. Two of them are shown in Figures \ref{fig:odd.scenario} and \ref{fig:even.scenario}. Each one depicts a shearless breakup in one of the reconnection scenarios of SNM.

Figure \ref{fig:odd.scenario} shows an odd reconnection scenario involving the shearless curve. Figure \ref{fig:odd.scenario}(a) displays the shearless curve (in red) located between twin chains of $19$ islands each (in cyan and magenta), evident in the magnification of rectangle area [Fig. \ref{fig:odd.scenario}(b)]. A slight decrease in the value of the control parameter $q_a$ leads to a collision of both island chains with the shearless curve and, in this process, the STB breaks up [Fig. \ref{fig:odd.scenario}(c)]. A further decrease in $q_a$ causes the reappearance of a shearless curve in a meandering form, separating the two island chains [Fig. \ref{fig:odd.scenario}(d)]. {Lowering further the parameter $q_a$,} this odd reconnection scenario involves the extinction of the twin islands, leaving only the meandering invariant curves, separated by the surviving shearless curve [Fig.  \ref{fig:odd.scenario}(e)]. 

A sequence of Poincaré sections showing an even reconnection scenario is depicted in Figure \ref{fig:even.scenario}. For a given value of $q_a$ there are twin chains of $4$ islands each (marked green and blue), straddling a shearless curve (in red) [a magnification of the phase portrait, showing only two islands in each chain, is displayed by Fig. \ref{fig:even.scenario}(a)]. For the current value of the amplitudes considered in this Figure, there are already two sizeable chaotic regions at both sides of the shearless curve, where the remnants of the islands are embedded. 

As the control parameter $q_a$ slightly increases, the twin island chains approach each other and reconnect, causing a transition when the shearless curve is absent [Fig.  \ref{fig:even.scenario}(b)], although the shearless transport barrier is still present. Increasing again $q_a$, the lower island chain (green) disappears, leaving only the upper chain (blue), showing an asymmetry characteristic of this even reconnection scenario [Fig. \ref{fig:even.scenario}(c)]. After this remaining island disappears, for increasing $q_a$, the shearless curve reappears [Fig. \ref{fig:even.scenario}(d)]. The asymmetric behaviour of this even reconnection scenario is a characteristic of nontwist maps with lack of symmetry, such as the considered model. For example, the standard nontwist map, Eq. \eqref{snm1}, has a symmetry with respect to action coordinate \cite{diego2}. However, most other maps do not have such property, like the extended standard nontwist map \cite{wurm2013}.

\begin{figure}
\centering
\includegraphics[width=0.9\textwidth]{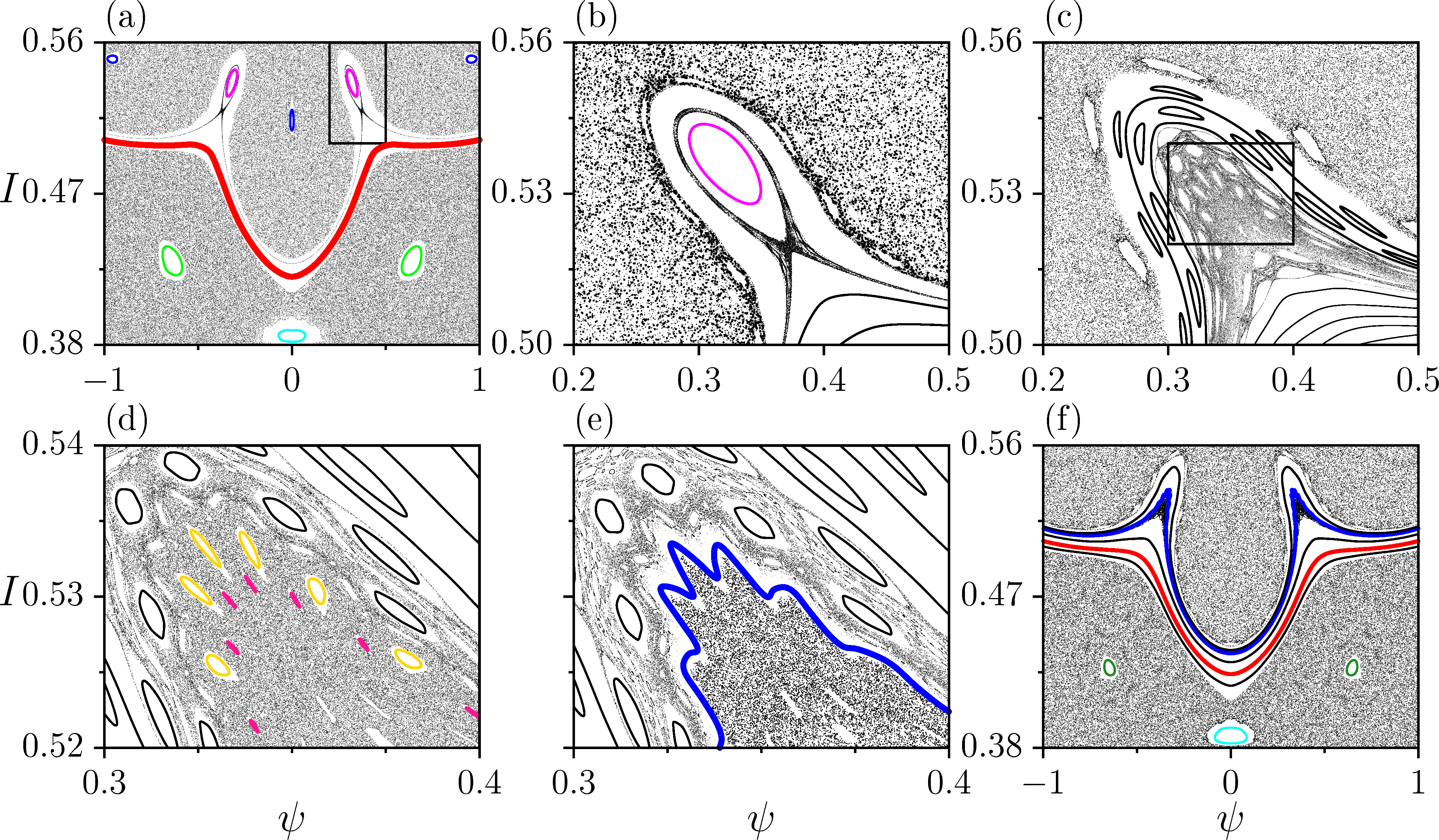}
\caption{\label{fig:poincare.bifurcation1} Poincaré sections of the drift wave model for (a) $q_a=3.2360$, (b) magnification of a region of (a), (c) $q_a=3.23841$, (d) magnification of a region of $q_a=3.23841$, (e) $q_a=3.23849$ {and (f) is a zoom out of (e)}. The remaining parameter values are the same as in Figure 2. This sequence represents a scenario containing four isochronous chains (magenta, cyan, blue, and green) originating shearless bifurcations. The periodic points of the magenta chain, in (a) and (b), collide in a saddle-center bifurcation (c) and a second shearless curve arises in (d) after the suppression of pink and gold twin island chains, containing 73 islands each (e)}.
\end{figure}

Another category of shearless bifurcations concerns the appearance of more than one shearless curve, as illustrated by Fig. \ref{fig:poincare.bifurcation1}. For $q_a = 3.2360$ there are four island chains, two of them at each side of a shearless curve, marked in magenta, blue, green, and cyan. As in the previous Figure, the amplitude of the perturbation is already large enough to create two large chaotic regions at each side of the shearless curve, which act as a transport barrier in this case, preventing large scale excursion of particles in chaotic orbits. 

Let us consider the vicinity of one island of the magenta chain in more detail, as revealed by the magnification shown in Fig. \ref{fig:poincare.bifurcation1}(b). Changing the control parameter value, the periodic points of the chain collide and a chaotic layer takes their place, containing a myriad of islands [Fig. \ref{fig:poincare.bifurcation1}(c)]. This collision represents a saddle-center bifurcation: we have pairs of periodic points, half of them locally stable (centres) and the other half unstable (saddles). As the control parameter increases, these pairs of periodic points approach each other and eventually coalesce at the bifurcation points, disappearing afterwards, and leaving a chaotic layer  therein. In this myriad of islands there is a pair of twin island chains, with 73 islands each marked in gold and pink, having the same rotation number, [Fig. \ref{fig:poincare.bifurcation1}(d)]. A bifurcation process extinguishes those twin island chains and generates a second shearless curve, plotted in blue in Fig. \ref{fig:poincare.bifurcation1}(e). {Finally, Figure \ref{fig:poincare.bifurcation1}(f) displays the full scenario with the two shearless curves.} We remark that the appearance of the second shearless curve is a bifurcation in the rotation number profile, whereas the saddle-center (O-X) bifurcation which precedes the latter occurs in the phase space (here represented by Poincaré sections). Bifurcations of this type have been observed in some atypical scenarios in the SNM \cite{wurm2005meanders}. 

\begin{figure}
\centering
\includegraphics[width=0.9\textwidth]{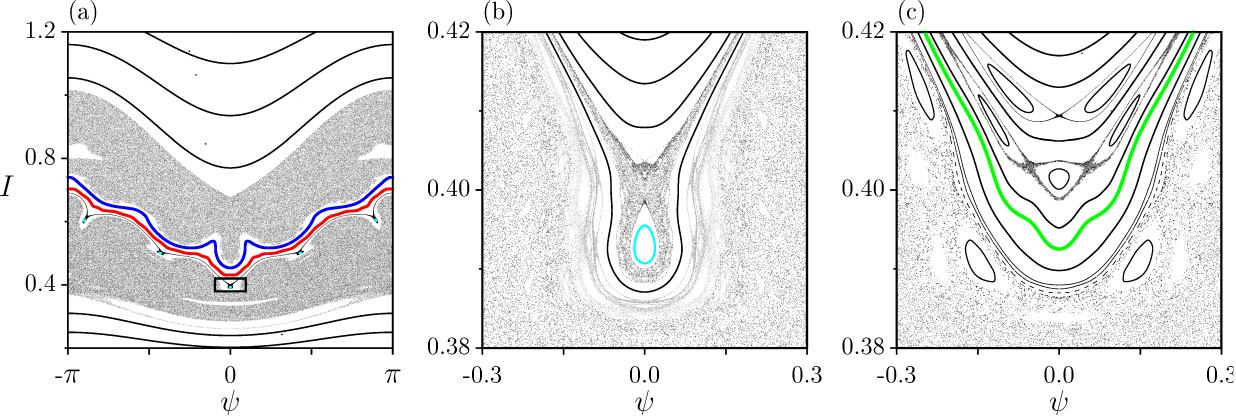}
\caption{\label{fig:poincare.bifurcation2} Poincaré sections of the drift wave model for (a) $q_a=3.2451$, (b) magnification of a region of (a), (c) $q_a=3.2527$. The remaining parameter values are the same as in Figure 2. This scenario illustrates the emerging of a third shearless curve through a saddle-node bifurcation.}
\end{figure}

The process detailed in Figure \ref{fig:poincare.bifurcation2} is analogous to Figure \ref{fig:poincare.bifurcation1} and leads to a third shearless curve, through the same kind of shearless bifurcation. In Figure \ref{fig:poincare.bifurcation2}(a), obtained for $q_a = 3.2451$, we show a scenario just after the appearance of a second shearless curve, straddling two chaotic regions. In this Poincaré section, below the shearless curve there is a chain of five tiny islands, one of them is magnified and shown in cyan in Fig. \ref{fig:poincare.bifurcation2}(b). Starting from this island (as well as the other ones in the same chain) a third shearless curve emerges, which is plotted in green in Fig. \ref{fig:poincare.bifurcation2}(c). In this example, another saddle-center bifurcation occurs in the Poincaré section, leading to a secondary twin islands chains with 58 islands each (not showed in the picture), which precedes the shearless bifurcation causing the emergence of the third shearless curve. The scenario after the emergence of the green shearless curve is depicted in Figure \ref{fig:poincare.general}(e).

\begin{figure}
\centering
\includegraphics[width=.8\textwidth]{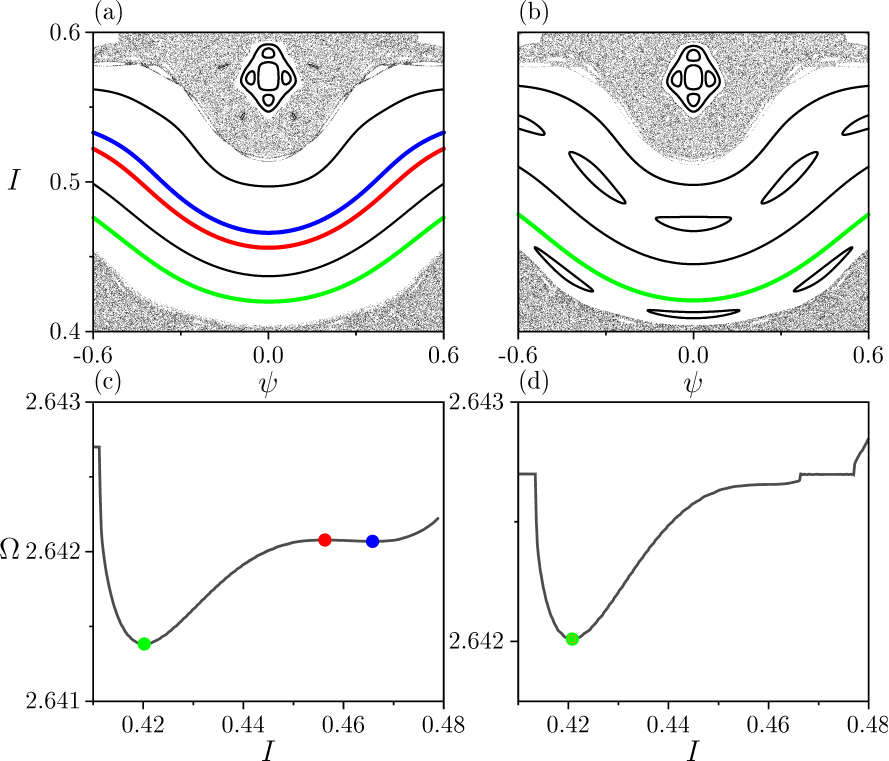}
\caption{\label{fig:poincare.bifurcation3} Poincaré sections of the drift wave model: (a) magnification of a region when $q_a=3.365$, (b) magnification of a region when $q_a = 3.367$. Rotation number profile for (c) $q_a=3.365$ and (d) $q_a = 3.367$. The remaining parameter values are the same as in Figure 2. This scenario exemplifies a bifurcation of shearless curves by collisions of periodic points (saddle-node). The red and blue curves collide and mutually annihilate.}
\end{figure}

The last scenario of shearless bifurcation presented in this paper is shown in Figure \ref{fig:poincare.bifurcation3}. For $q_a=3.365$ we see, in the Poincaré section, three shearless curves, marked in blue, red, and green [Fig. \ref{fig:poincare.bifurcation3}(a)]. The rotation number profile corresponding to this case [Fig. \ref{fig:poincare.bifurcation3}(c)] has three extrema: one maximum, corresponding to the red shearless curve, and two minima, corresponding to the green and blue curves. As the control parameter increases slightly, the red and blue shearless curves collide at a critical value of $q_a$. After the collision, these shearless curves mutually annihilate, leaving only the green shearless curve [Fig. \ref{fig:poincare.bifurcation3}(b)], as confirmed by the rotation number profile for this value of $q_a$ [Fig. \ref{fig:poincare.bifurcation3}(d)], which exhibits only the minimum corresponding to the green shearless curve.

\FloatBarrier
\section{Conclusions}

Shearless transport barriers have been identified in a theoretical model describing the $\mathbf{E}\times\mathbf{B}$ drift motion of the guiding centres. These barriers may be one of the mechanisms that hide behind certain improved scenarios observed in tokamaks with nonmonotonic plasma profiles. Three profiles have been specified in advance, following experimental evidences: the safety factor radial profile, the average radial electric field profile, and the toroidal velocity profiles. Excepting toroidal velocity, those profiles are nonmonotonic and thus present some kind of reversed shear, the combined effect of them being the key ingredient in our model. 

The equilibrium configuration characterized by these nonmonotonic profiles is perturbed by the influence of electrostatic fluctuations with a dominant mode, whose amplitude was also determined from experimental results in TCABR tokamak. Our results follow by numerical integration of the drift motion equations taking into account the reversed shear profiles as well as the electrostatic fluctuation field. Since the motion equations were expressed in action-angle variables, the drift motion has a Hamiltonian form, and the Poincaré sections (taken at integer multiples of the dominant mode of the perturbation period) are actually area-preserving maps of a non-integrable system.

Since the radial profiles used are nonmonotonic, this is a nontwist system, hence some novel features are expected with respect to models using monotonic profiles. One of them is the existence of shearless curves, located at extrema of the rotation number profiles (in action space, which is essentially the radial direction). 

The main result of our work was the discovery of multiple shearless curves, when one of the system parameters (namely the safety factor at the plasma edge) is varied in specified intervals. These intervals comprise the neighbourhood of $q_a = 3$ equilibrium magnetic surface, and thus are of physical interest, since many tokamaks (including TCABR) operate at that range. Such multiple shearless curves can appear and disappear due to shearless bifurcations, since they occur after (or before) the control parameter passes through critical values.

We identified three groups of shearless bifurcations. In the first group, we have shearless curves that simply breakup and reappear, indicating local changes of the rotation number profile. {In the second bifurcation group, new shearless curves appear in phase space, after a saddle-center bifurcation and emergence of secondary twin island chains. These curves are related to extrema of rotation number profile, but one of them corresponds to a local minimum, while the other one to a local maximum. Finally, the third group involves the collision and disappearance of shearless curves, due to a bifurcation in the rotation number profile, characterized by a maximum and a minimum points colliding as the control parameter is varied.} 

Depending on the strength of the fluctuation amplitudes, Poincaré sections show the existence of chaotic orbits on both sides of the shearless curves, indicating a limited extension of the chaotic transport in this region. In this case, when a shearless barrier breaks up, the chaotic orbits often merge together, leading to global transport. If there are more than one shearless curves, however, we can have two regions where the chaotic transport is reduced, increasing the complexity of the description. Since multiple internal transport barriers have been actually verified experimentally, our work sheds some light on the nature of this dynamical phenomenon. 

\

\noindent \textbf{Acknowledgments}

The authors thank the financial support from the Brazilian Federal Agencies (CNPq) under Grant Nos. 407299/2018-1, 302665/2017-0, 403120/2021-7, and 301019/2019-3; the São Paulo Research Foundation (FAPESP, Brazil) under Grant Nos. 2018/03211-6 and 2022/04251-7; and support from Coordenação de Aperfeiçoamento de Pessoal de Nível Superior (CAPES) under Grants No. 88887.522886/2020-00, 88881.143103/2017-01 and Comité Français d’Evaluation de la Coopération Universitaire et Scientifique avec le Brésil (COFECUB) under Grant No. 40273QA-Ph908/18.

\FloatBarrier

\appendix

\section{Normalization of variables in the drift-wave model}
\label{appendixA}
In this Appendix we outline the normalization of variables of the drift-wave model used in this paper, namely $B$, $a$, $\omega_0$, $\overline{E_r}$, $\phi_n$, and $v_{\|}$. SI units are used throughout. The minor plasma radius, $a$, is divided by the factor $a_0=0.18$ cm, so the normalized value $a'=a/a_0 = 1$. The same is done to the toroidal field: we choose $B_0=1.1$T $\Rightarrow$ $B'=B/B_0=1$. We adopt an electric field normalization factor in order that its magnitude has unit absolute value at the plasma edge: $E_0 = \left|\overline{E_r}(r=a)\right|$.
\begin{equation}
\overline{E_r} = 3\alpha r^2 + 2\beta r + \gamma,    
\end{equation}
\noindent where $(\alpha,\beta,\gamma) = (-80.00, 31.95, -6.00)\times 10^{3} \ \textrm{V/m}$. Thus, $E_0=2274 \ \textrm{V/m}$ and the normalized field, in action variables, is:
\begin{equation*}
\overline{E_r}' = \dfrac{\overline{E_r}}{E_0} = 3\dfrac{\alpha a^2}{E_0}\left(\dfrac{r}{a}\right)^2 + 2\dfrac{\beta a}{E_0}\left(\dfrac{r}{a}\right) + \dfrac{\gamma}{E_0},
\end{equation*}
\noindent therefore, 
\begin{equation}\label{eq:norm.electric.field}
\overline{E_r}' = 3\alpha' I + 2\beta' \sqrt{I} + \gamma',
\end{equation}
\noindent with $(\alpha',\beta',\gamma') = (-1.140, 2.529, -2.639)$.

The velocity, time, and electric potential normalization factors are given by: $v_0 = E_0/B_0 = 2067$ m/s, $t_o = a_0/v_0 = 8.707\times 10^{-5}$ s, $\phi_0 = a_0E_0 = 409.3$ V, respectively, whereas the lowest frequency is $\omega_0=6\times 10^4$ rads/s, such that its normalized value is given by $\omega'_0 = \omega_0 t_0 = 5.224$.

\printbibliography
\end{document}